\documentclass[10pt,conference]{IEEEtran}
\IEEEoverridecommandlockouts
\usepackage[ruled,linesnumbered]{algorithm2e}
\usepackage{multirow}
\usepackage[caption=false,font=normalsize,labelfont=sf,textfont=sf]{subfig}
\usepackage{listings}
\usepackage{booktabs}
\usepackage{multicol}
\usepackage{tabularx}
\usepackage{adjustbox}
\usepackage{makecell}
\usepackage{pifont}
\usepackage{enumitem}
\usepackage{calligra}
\usepackage{mdframed}
\usepackage{fontawesome}
\usepackage{amsmath,amssymb,amsfonts}
\usepackage{graphicx}
\usepackage{textcomp}
\usepackage{xcolor}
\usepackage{balance}
\usepackage{caption}

\def\BibTeX{{\rm B\kern-.05em{\sc i\kern-.025em b}\kern-.08em
    T\kern-.1667em\lower.7ex\hbox{E}\kern-.125emX}}
\lstdefinestyle{codeStyle}{
    basicstyle=\small\ttfamily,
    commentstyle=\color{green},
    keywordstyle=\color{blue},
    stringstyle=\color{blue},
    showstringspaces=false,
    breaklines=true,
    frame=lines,
    backgroundcolor=\color{white},
    captionpos=b,
    aboveskip=10pt,
    belowskip=10pt
}
\usepackage[most]{tcolorbox}

\definecolor{codebackground}{RGB}{240,240,240} % 

\newcommand{\appname}{{\sc VulEUT}\xspace}
\begin{document}
\title{Vulnerability-Triggering Test Case Generation from Third-Party Libraries}

% \textit{The State Key Laboratory of Blockchain and Data Security} \\

\author{\IEEEauthorblockN{Yi Gao\IEEEauthorrefmark{1}, Xing Hu\IEEEauthorrefmark{1}\IEEEauthorrefmark{3}\thanks{\IEEEauthorrefmark{3}Corresponding author.}, Zirui Chen\IEEEauthorrefmark{1}, Tongtong Xu\IEEEauthorrefmark{2}, Xiaohu Yang\IEEEauthorrefmark{1}}
\IEEEauthorblockA{\IEEEauthorrefmark{1}State Key Laboratory of Blockchain and Data Security, Zhejiang University\\
\IEEEauthorrefmark{2}Nanjing University\\
\{gaoyi01,xinghu,chenzirui,yangxh\}@zju.edu.cn, xttluck@gmail.com}
}

% \author{\IEEEauthorblockN{Yi Gao}
% \IEEEauthorblockA{
% \textit{Zhejiang University}\\
% Hangzhou, China \\
% gaoyi01@zju.edu.cn}
% \and
% \IEEEauthorblockN{Xing Hu}
% \IEEEauthorblockA{
% \textit{Zhejiang University}\\
% Hangzhou, China \\
% xinghu@zju.edu.cn}
% \and
% \IEEEauthorblockN{Zirui Chen}
% \IEEEauthorblockA{
% \textit{Zhejiang University}\\
% Hangzhou, China \\
% chenzirui@zju.edu.cn}
% \and
% \IEEEauthorblockN{Tongtong Xu}
% \IEEEauthorblockA{
% \textit{Nanjing University}\\
% Nanjing, China \\
% xttluck@gmail.com}
% \and
% \IEEEauthorblockN{Xiaohu Yang}
% \IEEEauthorblockA{
% \textit{Zhejiang University}\\
% Hangzhou, China \\
% yangxh@zju.edu.cn}

% Yi Gao (Zhejiang University) <gaoyi01@zju.edu.cn>
% Xing Hu (Zhejiang University) <xinghu@zju.edu.cn>
% Zirui Chen (Zhejiang University) <chenzirui@zju.edu.cn>
% Tongtong Xu (Nanjing University) <xttluck@gmail.com>
% Xiaohu Yang (Zhejiang University) <yangxh@zju.edu.cn>

% \thanks{Yi Gao, Xing Hu, Zirui Chen and Xiaohu Yang are with the State Key Laboratory of Blockchain and Data Security, Zhejiang University, Hangzhou, Zhejiang, China. E-mail: gaoyi01@zju.edu.cn, xinghu@zju.edu.cn, chenzirui@zju.edu.cn, yangxh@zju.edu.cn. Xin Xia is with the Zhejiang University, Hangzhou, Zhejiang, China. E-mail: xin.xia@acm.org. Xing Hu is the corresponding author.} 
% }

% \markboth{Journal of \LaTeX\ Class Files,~Vol.~18, No.~9, September~2020}%
% {Vulnerability-Triggering Test Case Generation from Third-Party Libraries}

\maketitle

\begin{abstract}
Open-source third-party libraries are widely used in  software development. 
These libraries offer substantial advantages in terms of time and resource savings. 
However, a significant concern arises due to the publicly disclosed vulnerabilities within these libraries.
Existing automated vulnerability detection tools often suffer from false positives and fail to accurately assess the propagation of inputs capable of triggering vulnerabilities from client projects to vulnerable code in libraries.
In this paper, we propose a novel approach called \appname (\underline{Vul}nerability \underline{E}xploit \underline{U}nit \underline{T}est Generation), which combines vulnerability exploitation reachability analysis and LLM-based unit test generation.
\appname is designed to automatically verify the exploitability of vulnerabilities in third-party libraries commonly used in Java client software projects.
\appname first analyzes the client projects to determine the reachability of vulnerability conditions. And then, it leverages the Large Language Model (LLM) to generate unit tests for vulnerability confirmation.
To evaluate the effectiveness of \appname, we collect 32 vulnerabilities from various third-party libraries and conduct experiments on 70 real client projects. 
Besides, we also compare our approach with two representative tools, i.e., TRANSFER and VESTA.
Our results demonstrate the effectiveness of \appname, with 229 out of 292 generated unit tests successfully confirming vulnerability exploitation across 70 client projects, which outperforms baselines by 24\%. 
\end{abstract}

\section{Introduction}
Modern software development highly depends on open-source third-party libraries (TPLs). 
Within the open-source community, TPLs play an important role in helping developers avoid redundant development efforts~\cite{lyu2023chronos,kula2018developers,yuan2022exploitability,samarinunderstanding}. 
Furthermore, the extensive and diverse libraries form a large-scale software ecosystem and provide developers with a wide range of tools, features, and functionalities~\cite{wang2020empirical,alfadel2023empirical,latendresse2022not}.
For example, the well-known Maven is a primary tool for managing the Java ecosystem, encompassing more than 9.51 million TPLs according to a recent report~\cite{apache-maven}. 
The presence of these libraries has remarkably facilitated the progress of Java project development.

Unfortunately, TPLs are prone to have vulnerabilities, and the number of publicly disclosed vulnerabilities in these libraries has been increasing~\cite{wu2023understanding}. 
These vulnerabilities pose significant security threats to the entire software ecosystem, particularly to all the projects dependent on vulnerable 
particularly for client projects depending on vulnerable TPLs~\cite{chen2019gui,tang2019large}. 
Zhan et al.~\cite{zhan2021atvhunter} point out that approximately 74.95\% of libraries with vulnerabilities are still widely used by their client projects. 
A recent notable example is the vulnerability in the Apache Log4j2 TPL, which affects over 35,000 Java packages and impacts millions of devices~\cite{wetter2021understanding,kang2022test,wu2023understanding,huang2022characterizing}.

The impact of vulnerabilities in TPLs has attracted increasing attention from both academia and industry. 
In recent years, automated tools have been introduced to detect and evaluate whether client projects are affected by vulnerabilities within the TPLs they depend on. 
For example, dependency-based approaches~\cite{alfadel2023empirical,decan2018impact,zapata2018towards,wu2023understanding} analyze dependency configuration files or gather compilation information through Software Composition Analysis techniques to identify vulnerable dependencies that may introduce security risks. 
Call graph-based approaches~\cite{nielsen2021modular,ponta2018beyond,foo2019dynamics,iannone2021toward,ponta2018beyond} assess if projects invoke vulnerable code within TPLs, primarily focusing on the invocation of methods within the vulnerable code.  

Recently, \textsc{TRANSFER}~\cite{kang2022test} collects and reuses existing vulnerability exploitation tests within TPLs, slicing these tests to capture program states relevant to triggering vulnerabilities.
However, due to the infrequent presence of existing vulnerability-triggering tests within TPLs~\cite{chen2023exploiting}, \textsc{TRANSFER} fail to be effective when applied to other TPLs.
\textsc{VESTA}~\cite{chen2023exploiting} ensures the similarity between generated tests and exploits by migrating parameters using specified transformation rules. 
However, this approach relies on manually crafted rules for test manipulation, making it ineffective in cases where the rules are not comprehensive.

Additionally, addressing vulnerabilities in client projects, such as updating TPLs and transferring to secure versions, may introduce other dependency conflicts or compatibility issues~\cite{xavier2017historical,huang2021repfinder,liu2021identifying,wang2019could}. 
Developers are frequently confronted with numerous inaccurate false positives due to the lack of measuring whether client projects can generate inputs to exploit vulnerabilities~\cite{chen2023exploiting}.
Consequently, there is a substantial need to propose more precise and automated tools capable of generating comprehensive tests to assess the security threats posed by vulnerabilities.

To generate unit tests capable of confirming the trigger ability of vulnerabilities in client projects, we need to address the following two challenges:

\noindent \textbf{Challenge 1: How to conduct fine-grained call path analysis?} 
While the call graph provides insights into the calling relationships between the client project and the vulnerable code, triggering a vulnerability within the TPL in a real client project necessitates the satisfaction of specific constraints. 
These constraints involve aspects such as the origin of parameters, the path of parameter propagation, and variations in parameter values.
We should conduct a fine-grained call path analysis to confirm whether the conditions required to trigger vulnerabilities can be effectively transferred from the client project's user access interface to the vulnerable code.

\noindent \textbf{Challenge 2: How to generate unit tests to confirm vulnerability exploitation?} 
The second challenge involves generating unit tests within the client project to confirm the vulnerability's exploitation and ascertain if it can be triggered as intended.
The unit test emulates an attacker's behavior by executing the program's user access interface and transferring the malicious inputs to the vulnerable code within the client, thereby triggering the vulnerability. 
Furthermore, the unit test is designed to confirm whether the vulnerability is successfully triggered automatically.

Recently, large language models (LLMs) such as ChatGPT~\cite{chatgpt} have demonstrated outstanding performance in both code comprehension and code generation tasks~\cite{lemieux2023codamosa,bareiss2022code,yuan2023no,schafer2023adaptive}.
In this paper, we propose to combine fine-grained call path analysis with LLM-based test generation techniques to investigate the triggerability of TPL vulnerabilities within Java client projects.

Our proposed technique, named \appname, first analyzes the source code of the client project to obtain the vulnerability's reachability within the client project. By analyzing the method call path to assess the reachability from the user access interface to the vulnerable code of the TPL, we take a step further by constructing a Parameter Transfer Graph (PTG), to identify the origins of vulnerability conditions within the client project. 
Subsequently, in coordination with predefined parameter transfer rules, we confirm the reachability of vulnerability conditions between the client project and the TPL.
Next, \appname utilizes code context extraction and ChatGPT's test generation capability to produce unit tests for exploiting vulnerabilities.
Finally, we run the generated test files in the client project and explore the triggerability of vulnerabilities.
% We begin with a vulnerability reachability analysis of the client project. 
% By analyzing the method call path to assess the reachability from the user access interface to the vulnerable code of the TPL, we take a step further by constructing a Parameter Transfer Graph (PTG).

% We design the PTG to identify the origins of vulnerability conditions within the client project.
% Subsequently, in coordination with predefined parameter transfer rules, we confirm the reachability of vulnerability conditions between the client project and the vulnerable code of the TPL.      
% Next, we perform vulnerability triggering confirmation for the client project. 
% Based on the preceding analysis, we extract code contexts and leverage ChatGPT's test generation capability to produce unit tests for exploiting vulnerabilities. 

We evaluate the performance of \appname by collecting publicly disclosed vulnerabilities from 20 different types of TPLs and obtaining 70 real client projects from GitHub. 
In our experiments, we generate a total of 292 unit tests, with 229 of them effectively confirming vulnerability exploitation. 
Furthermore, among the 70 projects, 56 received successful confirmation for vulnerability exploitability, which outperforms baselines by 24\%.

The main contributions of this paper are as follows:
\begin{itemize}[leftmargin=*]
\item We propose a novel approach that combines fine-grained call path analysis and the testing capabilities of LLMs to substantiate the exploitation of vulnerabilities from TPLs within client projects, thereby reducing false positives.
\item We implement \appname, a tool capable of generating unit tests for client projects to automatically confirm whether TPL vulnerabilities are triggered. 
Our tool is available on our website~\cite{vulEUT}.
\item Experiments with 20 TPLs, 35 vulnerabilities, and 70 open-source projects demonstrate the effectiveness of the \appname. 
It outperforms the baseline by achieving a 24.44\% increase in successfully confirmed vulnerability-triggering tests.
\end{itemize}

% \begin{figure}
%     \begin{lstlisting}[language=Java,xleftmargin=1em]
% @Test public void testXml2ObjCallsVulnerabilityMethod() {
%     String input = "<void>";
%     // Set up an interceptor to detect calls to the XStream.fromXML method
%     MethodCallInterceptor.interceptor(
%         com.thoughtworks.xstream.XStream.class, "fromXML", new Object[]{input});
%     try {
%         XmlUtil.xml2Obj(input, Object.class);
%         fail("Expected Exception");
%     } catch (Exception e) {
%         e.printStackTrace();
%     }
%     // Verify that the vulnerability is successfully triggered
%     assertTrue(MethodCallInterceptor.isTriggered());
%     assertTrue(MethodCallInterceptor.isConditionMet());
% }
%     \end{lstlisting}
%   \caption{An example of vulnerability triggering unit test generated by VulEUT.}
% \label{fig:motivation-example}
% \end{figure}
\begin{figure*}
\centerline{\includegraphics[width=0.95\textwidth]{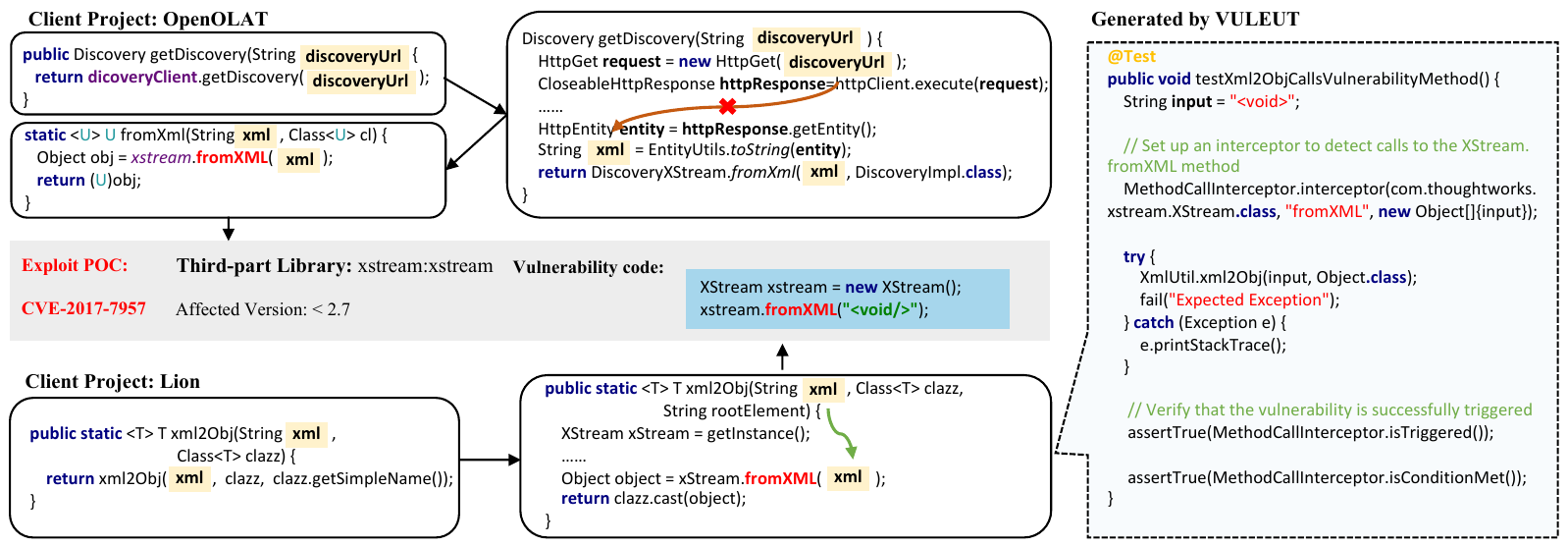}}
% \vspace{-0.1cm}
\caption{An example of TPL vulnerability exploitation.}
\label{fig:tpl-exploitation}
\vspace{-0.4cm}
\end{figure*}

\section{Motivation}
\label{sec:motivation}

While vulnerabilities disclosed through CVEs allow developers to identify dependencies on vulnerable versions of TPLs in their client projects, the existence of a package dependency on TPLs does not ensure the triggering of vulnerabilities inside these projects.
For example, even if a client project depends on a TPL with known vulnerabilities, it does not indicate the utilization of the specific vulnerable methods within the library.
Methods based on call-graph analysis help to confirm whether client project methods call TPL vulnerable methods.
Although a call relationship exists between methods of the client project and vulnerable methods, it does not conclusively prove a security threat. 
Further investigation is required to confirm whether external users can exploit this vulnerability through the client projects’ user access interface, which serves as an accessible interface under specific input conditions.

For example, in Figure~\ref{fig:tpl-exploitation}, \textit{XStream} is primarily used for Java object-to-XML conversion, applied to data persistence and cross-platform data exchange, and other scenarios.
CVE-2017-7957 exposed a Denial-of-Service (DoS) vulnerability within the \textit{XStream} library's \texttt{fromXML} method.
we present two client projects, both of which depend on vulnerable versions of \textit{XStream} ($<= 1.4.9$). 
After manually inspecting the vulnerable code, we find that the client project \textit{OpenOLAT} exhibits a call relationship on the \textit{XStream} \texttt{fromXML} method within its own \texttt{fromXML} method. 
In contrast, the client project \textit{Lion} has it within its \texttt{xml2Obj} method. 
It is important to note that only the existence of a method call relationship is insufficient to exploit the vulnerability.
Therefore, we should further analyze the propagation pattern of vulnerability conditions (e.g., method parameters) during method execution, to confirm whether trigger conditions can traverse from the user access interface to the vulnerable method.

Specifically, we first examine the \texttt{fromXML} method to investigate its calling relationships within the \textit{OpenOLAT}. 
Remarkably, it is established that only the \texttt{getDiscovery} method invokes it.
Then, we further examine the parameter propagation pattern. 
We find that the entry method, \texttt{getDiscovery}, has only one parameter, \texttt{discoveryUrl}, which is used to initiate a network request.
Following this, the outcome of the network request is parsed into an \texttt{xml} parameter and is eventually passed to the vulnerable method, \texttt{fromXML}. 
Due to the necessity of specially-crafted XML containing malicious characters to exploit the vulnerability, and considering the significant functional and propagation differences between the variables \texttt{discoveryUrl} and \texttt{xml}, the propagation of malicious input from the user access interface to the vulnerable method is unreachable.
In contrast, within the \textit{Lion} project, the \texttt{xml2Obj} method's \texttt{xml} parameter is directly passed to the vulnerable \texttt{fromXML} method through the method call path. 
Consequently, vulnerability-triggering conditions can directly access the vulnerable code in this case, placing it in a state of security risk.

To further automate the confirmation of vulnerability triggerability, our approach, \appname, generates a unit test for entry method associated with call path, as shown in Figure~\ref{fig:tpl-exploitation}. 
This test includes the following components: The entry method of the method call path serves as the \textit{focal method} and is specifically denoted as \texttt{xml2Obj}. 
The \textit{test input} encompasses the disclosed vulnerability input conditions, which serves as input data for the unit test.
\texttt{MethodCallInterceptor.interceptor} is a generic vulnerability detector used to check whether TPL vulnerability code is triggered during the test execution and whether vulnerability trigger conditions are met.

Regarding the \textit{test oracle}, the primary purpose of this unit test is to confirm whether the TPL vulnerability affects the client project. 
We design the test oracle in three parts: 
(1) assert whether the vulnerability code execution is achievable through the user access interface (\texttt{isTriggered()}); 
(2) to assert whether the vulnerability trigger conditions are met during the execution of the vulnerability code (\texttt{isConditionMet()}); 
(3) to assert whether the program behavior meets the expectations when the vulnerability is triggered based on its type (\texttt{fail("Expected Exception")}). 
If the unit test passes, it indicates that the client project is currently at a disclosed vulnerability exploitation risk, prompting developers to address the vulnerability.

\section{approach}
\label{sec:approach}

\begin{figure*}
\centering
\includegraphics[width=0.95\linewidth]{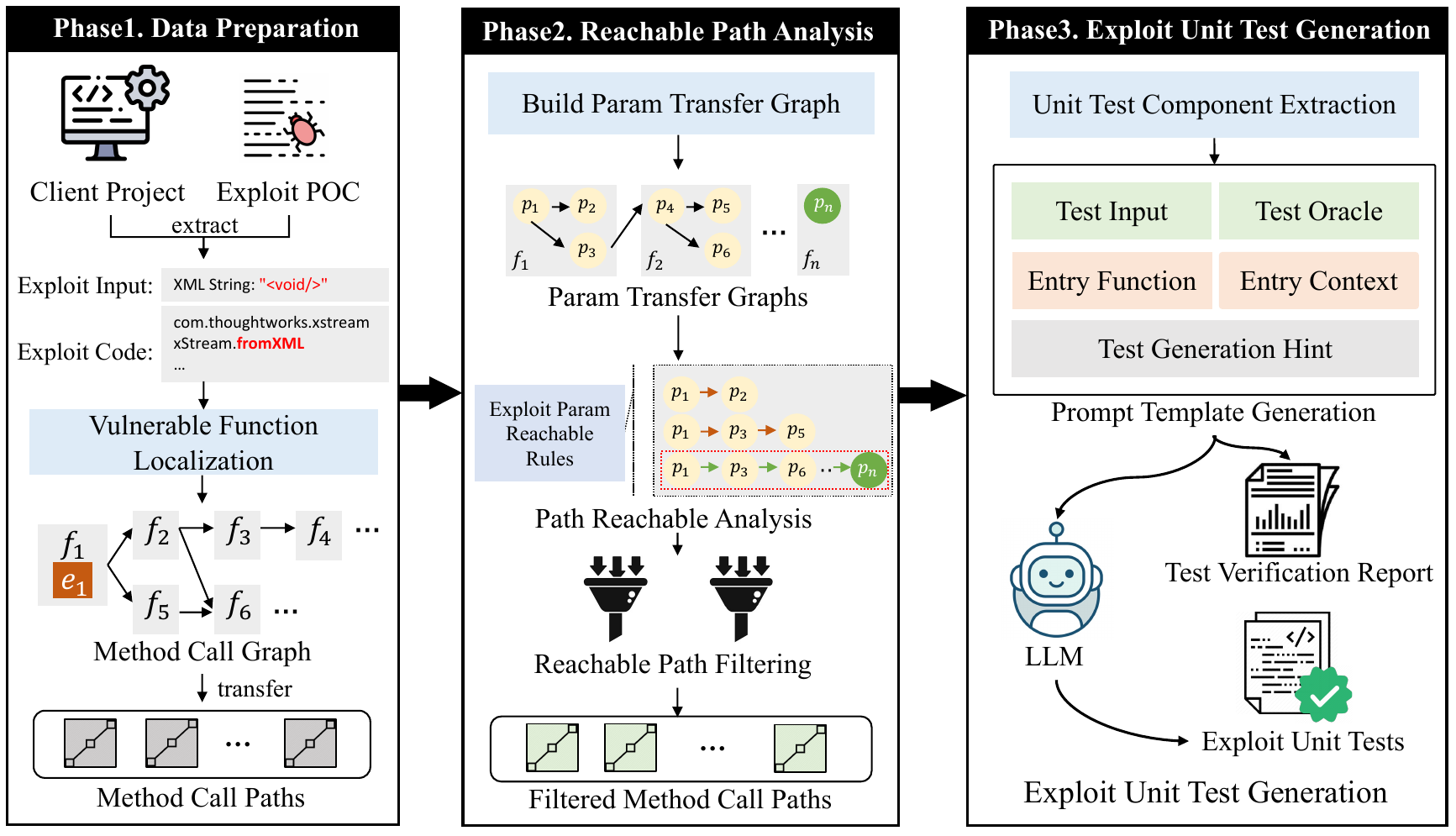}
\caption{Overview of our Approach.}
\vspace{-0.3cm}
\label{fig:approach}

\end{figure*}

This section introduces the details of our proposed approach \appname. Figure~\ref{fig:approach} illustrates the overall framework of \appname, which can be divided into three main phases:

\textbf{Phase} \ding{182} \textbf{Data Preparation.}
During the data preparation phase, we initially extract the source code structure of the client project and code information regarding vulnerabilities in TPL. 
Additionally, we identify vulnerable methods within the client project where the vulnerable code is utilized.
Subsequently, we construct the method call graph associated with the vulnerable code. 

\textbf{Phase} \ding{183} \textbf{Reachable Path Analysis.} 
In this phase, the primary goal is to perform a reachability analysis of vulnerabilities. 
This involves analyzing the path within the target client project to determine whether the conditions for exploiting a vulnerability can propagate from the client's user access interface to the vulnerable code within the TPL.

\textbf{Phase} \ding{184} \textbf{Exploit Unit Test Generation.} 
This phase confirms whether we can trigger a TPL's vulnerability in the client project.  
We leverage the results obtained from the previous phase to identify cases where vulnerabilities are reachable, including method call paths, entry methods, parameter specifications, and vulnerability trigger conditions. 
Subsequently, this phase generates unit tests for the identified entry methods to trigger the vulnerabilities. 
Finally, we generate an exploitation confirmation report by executing the generated unit tests.

\subsection{Data Preparation}
As shown in Figure~\ref{fig:approach}, during this phase, we utilize the client project's source code and publicly disclosed vulnerability reports (i.e., PoC~\cite{poc}) as input. 
This phase primarily consists of three parts, i.e., information extraction, vulnerable method localization, and method call path construction.

\subsubsection{Information Extraction}
We extract structural information from the source code of the client project and details related to vulnerable code obtained from vulnerability reports. And then we identify the location of the vulnerable code within the client project's methods.
We utilize ANTLR~\cite{antlr} to extract code structures from the client project, including classes, methods, and fields, and construct their corresponding abstract syntax trees. ANTLR provides a straightforward syntax rule language commonly employed for parsing tasks in various programming languages.
These extracted structures are uniformly stored using a standardized data structure, forming the basis for subsequent analysis.
Regarding vulnerability reports (e.g., POC), we extract segments of vulnerable code, encompassing vulnerable methods within TPLs, their corresponding classes, and triggering conditions.

\subsubsection{Vulnerable Method Localization}
We construct method call graphs within client methods. 
We aim to locate and extract client methods that establish calling relationships with specific vulnerable code within the TPL. 
Our approach involves iteratively examining invocation statements within each method and parsing them to obtain full class names and method signatures. 
These details are then compared with the segments of vulnerable code. 
If a match is found, the corresponding client method is marked as vulnerable.

\subsubsection{Method Call Path Construction}
% We reverse-traverse the call graph from the vulnerable methods to obtain all method call paths associated with the TPL's vulnerable code. 
% This serves as the starting point for vulnerability reachability analysis.
To obtain the method call paths in the client project that depend on vulnerable code from TPLs, we first utilize the Soot~\cite{soot} to construct a method call graph. 
Soot is an open-source Java program analysis and optimization framework and offers extensive static and dynamic analysis capabilities.
Soot’s default method call graph construction uses Class Hierarchy Analysis (CHA), but this can be inaccurate due to the lack of dynamic inheritance and polymorphism mechanisms in the Java. 
Therefore, we employ the Spark algorithm (\texttt{-p cg.spark}) in Soot to construct the call graph, which exhibits good performance. 
Additionally, we enable Soot's -\texttt{allow-phantom-refs} option to handle phantom references, including reflective calls. 
This involves collecting potential targets of reflective calls to construct the call graph. 
The Spark algorithm in Soot further deals with polymorphism by considering type information and method call context. 
When constructing the call graph, we specify a single API within the TPL as the starting point and search internally within the client project, significantly reducing the search space. 
Furthermore, we set filtering conditions, such as discarding paths related to testing (e.g., \texttt{@Test}) or paths with restricted access permissions (e.g., \texttt{private}), further reducing the space of paths.

\subsection{Reachable Path Analysis}
Although method call paths indicate the existence of call relationships from the program's user access interface to the vulnerability method, 
the lack of vulnerability condition assessment does not ensure the successful exploitation of the vulnerability in the client program. 
Therefore, we explore whether the vulnerability code can be executed under specific conditions that trigger the vulnerability, like crafted malicious parameter values. 
We conduct a more fine-grained analysis to investigate the propagation of vulnerability conditions through method call paths.

This phase integrates the extracted code structure and method call paths from last phase to construct a parameter transfer graph (PTG) to analyze whether the triggering parameters of the vulnerability can be propagated from the program's user access interface to the specific location of the vulnerability code. 
Then, we gather information on accessible paths for the vulnerability conditions, which will be utilized to generate unit tests to confirm vulnerability exploitation.

\subsubsection{Parameter Transfer Graph Construction}
For each method with the method call path, we construct a Parameter Transfer Graph (PTG), facilitating the analysis of the transformation process of vulnerability parameters within the method. 
This graph determines whether vulnerability conditions can propagate from the method entry to the specific location of the vulnerability code within the method.
In terms of PTG structure, specific variables appearing within a method serve as nodes, while code statements (such as assignment statements and method call statements) act as edges. 
The PTG comprises data structures with the format $\langle$\textit{SourceNode}, \textit{TargetNode}, \textit{Edge}$\rangle$, where \textit{SourceNode} represents variables appearing on the left side of code statements (e.g., assignment statements), \textit{TargetNode} represents variables on the right side of assignment statements, and \textit{Edge} represents the code statements corresponding to the relationship between the two nodes. 
For example, a code statement \texttt{String xml = EntityUtils.toString(entity)} forms a tuple $\langle$ \texttt{entity}, \texttt{xml} , \texttt{EntityUtils.toString} $\rangle$, where \texttt{entity} is the source node, \texttt{xml} is the target node, and \texttt{EntityUtils.toString} serves as the edge representing the method call statement. 
Then, we leverage this graph to deduce the transfer paths for specific parameters, extracting patterns of parameter propagation.

\subsubsection{Path Reachable Analysis}
Here, we aim to analyze the parameters appearing in the vulnerability code within the client project, understand their propagation patterns, and determine their reachability.
We start with these parameters as the starting points, then trace their transformation process within the method call paths, working in an upward direction.

To identify parameter transformation patterns within a method, we formulate a set of rules outlining parameter transfer, categorized into four classes, as presented in Table~\ref{tab:rule-desc}.
In the first two categories, namely direct propagation and type conversion, the parameter value remains unchanged during propagation, enabling it to propagate downstream without modification.
In contrast, in cases involving Value Change and No Propagation, the parameter value is modified by intermediate processes, causing it unable to further propagate.
For example, in Figure~\ref{fig:tpl-exploitation} of the \textit{OpenOLAT} project, the input parameter \texttt{discoveryUrl} serves the purpose of initiating a network request.
However, the parameter \texttt{xml} that actually triggers the vulnerability in this example originates from the result of an internal network request.
Consequently, this parameter path is incapable of propagating malicious parameter values from the user access interface to the vulnerable code.

\begin{table}[h]
\centering
\caption{Vulnerability Parameter Propagation Rules.}
\begin{tabular}{l p{5.5cm}}
\toprule
\textbf{Propagation Rule} & \textbf{Description} \\
\midrule
Direct Propagation & 
The parameter is passed directly from the input to the called method without any changes. \\

\hline

Type Conversion & 
The parameter's type changes. For example, it may involve explicit type casting, transforming a field into an \texttt{Object}, or converting a \texttt{String} to an \texttt{Object}. \\

\hline

Value Change & 
During propagation, the parameter undergoes logical operations resulting in a different value. \\

\hline

No Propagation & 
It generates intermediate results, where the parameter itself does not propagate further. \\

\bottomrule
\end{tabular}
\label{tab:rule-desc}
% \vspace{-0.3cm}
\end{table}

\begin{algorithm}\small
\caption{Parameter Transfer Analysis}\label{algo:transfer}
\KwIn{$methodCallPath$}
\KwOut{$pTransferTypes$: the parameter transfer types for the method}

\BlankLine
\SetKwProg{Fn}{Function}{:}{}
\Fn{AnalyseParameterTransferType($methodCallPath$)}{
    $pTransferTypes \gets$ Initialize an empty list\;
        
    \ForEach{$method$ in $methodCallPath$ from end to start}{
        $graph \gets$ Initialize an empty graph\;
        $callerParams \gets$ getFromMethod($method$)\;
        $calleeParams \gets$ getFromCallee($method$)\;
        $statements \gets$ getStatements($method$)\;

        \ForEach{$cp$ in $calleeParams$}{
            % \tcp{for each callee param, construct a parameter transfer graph within the method.}
            $corStatements \gets$ locateStatements($cp$, $statements$)\;
            $graph \gets$ buildPTG($cp$, $corStatements$, \linebreak $callerParams$)\;
            
            $paramPaths \gets$ buildPaths($cp$, $graph$)\;
            \ForEach{$pp$ in $paramPaths$}{
                $transferType \gets$ analyseTransfer($corStatements$, $pp$)\;
                $pTransferTypes$.add($transferType$)\;
            }
        }
    }
    
    \Return{$pTransferTypes$}\;
}
\end{algorithm}

Algorithm~\ref{algo:transfer} outlines the entire process of analyzing parameter propagation patterns.
Specifically, we initiate this process with the last method in the method call path and create a PTG for each parameter within a single method (Lines 3-10).
After constructing the PTG for parameters within a method, we can extract the propagation patterns of these parameters within the method. 
To achieve this, we traverse the entire PTG starting from the parameters to obtain propagation paths (Line 11). 
It is important to note that there may be multiple paths, and we collect all of them. 
We establish a set of rules to analyze the propagation types along these paths (Line 13).
The definitions of these rules are as follows:

\begin{itemize}[leftmargin=*]
\item If a parameter undergoes no changes and directly originates from the method's input parameters, it is classified as \textit{Direct Propagation}.
\item If a parameter only undergoes type changes, such as from a string to an object, it is classified as \textit{Type Conversion}.
\item Code statements involving logical operations are categorized as \textit{Value Change}.
\item If a parameter's initial source within the method is not from the input parameters, it is classified as \textit{No Propagation}. 
This typically indicates an internal source, as shown in the \textit{OpenOLAT} example in Figure \ref{fig:tpl-exploitation}.
\end{itemize}

Our algorithm collects all the transfer types along the parameter path. 
If only \textit{Direct Propagation} or \textit{Type Conversion} appears in the path, the parameter is deemed reachable within the method.
However, if \textit{Value Change} or \textit{No Propagation} types are present, it is considered unreachable within the method. 
This process continues iteratively, determining the parameter's reachability along the method call path.

\subsection{Exploit Unit Test Generation}
Based on the aforementioned analysis results, we can evaluate the reachability of vulnerabilities. 
To automate the confirmation of vulnerability triggerability, we generate unit tests for the program's user access interface to trigger the vulnerabilities in the client project. 
This proves the project is exposed to security threats posed by TPL vulnerabilities.
During this phase, we leverage LLM, e.g.,  ChatGPT~\cite{chatgpt}, to generate unit tests automatically. 
ChatGPT exhibits remarkable performance in code comprehension and test generation~\cite{bareiss2022code}. 
We explore combining the results of reachability analysis with ChatGPT to automatically generate unit tests capable of confirming vulnerability triggers.

\subsubsection{Unit Test Component Extraction}
We extract the comprehensive test structure to build a unit test from the method call paths and generate specific prompts for ChatGPT. 
Given that the client project is implemented in Java, we configure the unit test framework to use JUnit~\cite{junit} for test generation.

Each unit test consists of a test prefix and a test oracle. The test prefix includes the preconditions and the focal method. 
We use the vulnerability trigger conditions extracted from the vulnerability report (such as parameter configurations) as input data for the unit test. 
Due to our aim to simulate attackers exploiting vulnerabilities in the client project, and the method at the beginning of the path serves as the accessible interface for users, we choose it as the focal method for the unit test.
Regarding the test oracle, since our designed unit test aims to confirm whether a vulnerability can be triggered in the client rather than validating the behavior of the focal method, we implement dual validation. 
First, during a single program execution, we check whether the vulnerable code in the TPL is executed. 
Second, we confirm whether the conditions for triggering the vulnerability are satisfied during its execution.

% To achieve this, we introduce the \texttt{MethodCallInterceptor}, a call interception tool implemented using the \textit{ByteBuddy} framework~\cite{buddy}. The main functionality of this class is to dynamically monitor whether a specified method in a given class is executed at runtime, intercept the method execution, and record variable values. The \texttt{isTriggered} method provided by \texttt{MethodCallInterceptor} returns a boolean value indicating whether the specified method is executed,\gao{}. 
% Therefore, we incorporate the \texttt{MethodCallInterceptor} and design the test oracle to rely on whether the \texttt{isTrigger} method returns true. This signifies that the specified vulnerability code is triggered under the specified conditions.
To achieve this, we introduce the \texttt{MethodCall Interceptor}, a call interception tool implemented using the \textit{ByteBuddy} framework~\cite{buddy}. 
Its main functionality is to dynamically monitor whether a specified method in a given class is executed at runtime, intercept the method execution, and record variable values. 
The \texttt{isTriggered} method provided by \texttt{MethodCallInterceptor} determines whether the specified method is executed, while the \textit{isConditionMet} method evaluates whether the conditions for triggering the vulnerability are satisfied by comparing parameter values during the execution. 
Consequently, we incorporate the \texttt{MethodCallInterceptor} into our design and construct the test oracle, signifying the triggering of specified vulnerability code under predetermined conditions.

\subsubsection{Unit Test Generation}

We use the example in Section~\ref{sec:motivation} to illustrate the process of constructing a prompt, as shown in Table~\ref{tab:prompt-template}.
The prompt encompasses test inputs, the focal method, and the test oracle.
This is a straightforward example, and in actual cases, vulnerability trigger conditions for TPL can be more intricate.
For instance, there may be multiple parameter types and additional trigger configuration conditions. We append these conditions to the prompt in a similar manner.
Furthermore, as method parameters may encompass reference types specific to the client project, we supply corresponding source codes to facilitate ChatGPT's understanding of their definitions.
Then, we transmit the prompts to ChatGPT through the provided API.
Our test configuration is notably explicit, offering comprehensive information regarding test inputs, the focal method, and the test oracle.
Due to potential parameter differences between vulnerability trigger conditions and entry method parameters, such as an additional \texttt{encoding} parameter in the entry method, tests generated by LLM can exhibit diversity in addressing multiple execution paths, for instance, by generating \texttt{UTF-8}, \texttt{ISO-8859}, and other parameter values.

\begin{table}[h]
\centering
\caption{Prompt Examples for Exploit Unit Test Generation.}
\label{tab:template}

\begin{tabular}{p{1.7cm} p{6cm}}
\toprule
\textbf{Component}& \textbf{Description} \\ 
\midrule

Prompt Hint &
\textbf{Role:} I want you to act like a Java tester.
\textbf{Hint:} Generate a \textbf{unit test} for confirming vulnerability exploitation using the \textbf{JUnit} framework, with the following requirements: \\

\hline
Focal Method &
\textbf{Hint:} The focal method is \textcolor{blue}{xml2Obj}, located in the \textcolor{blue}{XmlUtil} class,  the method signature is:
\textbf{Code:}
\begin{lstlisting}[frame=none, aboveskip=1pt, belowskip=-8pt]
public static <T> T xml2Obj(String xml, Class<T> clazz)
\end{lstlisting}
\\

\hline
Test Input & 
The \textbf{input variable name} for this unit test is \textcolor{blue}{input}, and the \textbf{value} is: \textcolor{red}{\texttt{<void>}};
\\
\hline
Test Oracle &
\textbf{Hint1:} The assert statement to confirm that the vulnerability is successfully triggered is fixed as:
\textbf{Code:}
\begin{lstlisting}[frame=none, numbers=none, belowskip=1pt]
assertTrue(MethodCallInterceptor.isTriggered());
assertTrue(MethodCallInterceptor.isConditionMet());
\end{lstlisting}
\textbf{Hint2:} The vulnerability type is \textit{Uncaught Exception}. After invoking the method, proceed with:
\begin{lstlisting}[frame=none, numbers=none, belowskip=-8pt]
fail("Expected Exception")
\end{lstlisting}
\\

\hline
Vulnerable Method  & 

\textbf{Hint:} The vulnerable method is \textcolor{blue}{xml2Obj}, located in the \textcolor{blue}{XmlUtil} class,  the method signature is:
\textbf{Code1:}
\begin{lstlisting}[frame=none, numbers=none, belowskip=2pt]
public static <T> T xml2Obj(String xml, Class<T> clazz, String rootElement) 
\end{lstlisting}

The vulnerable code snippet is:  
\textbf{Code2:}
\begin{lstlisting}[frame=none, numbers=none, belowskip=-8pt]
Object object = xStream.fromXML(xml);
\end{lstlisting}
\\

\bottomrule
\end{tabular}
\label{tab:prompt-template}
\vspace{-0.2cm}
\end{table}

\subsubsection{Test Confirmation}
In this phase, we undertake the compilation and execution of the previously generated unit tests. 
Initially, we gather these tests, relocate them to the designated test directory within the target client project, and then run them manually.
After that, we conduct a compilation check. 
If any unit test exhibits syntax errors or fails to compile, we record the error details and exclude the test.
For those tests that compile successfully, we proceed to execute them using the JUnit~\cite{junit} framework. 
This phase allows us to ascertain whether each test passes or fails, and we systematically document the execution outcomes in a comprehensive report.
In this way, the generated tests are not only syntactically correct but also functionally executable, effectively confirming the trigger conditions of the vulnerability.
\section{evaluation}
\label{sec:evaluation}
Our experiments are designed to address the following research questions:

\begin{itemize}[leftmargin=*]
\item \textbf{RQ1: How effective is VulEUT in confirming the trigger ability of vulnerabilities?}

% This question primarily focuses on evaluating \appname's effectiveness in vulnerability confirmation. 
% We collect and analyze the number and accuracy of vulnerability confirmation tests generated by the \appname. 
% We compare these results with those generated by \textsc{TRANSFER} and \textsc{VESTA}.

\item \textbf{RQ2: What is the impact of reachability analysis and prompt design on the effectiveness of VulEUT?}

We investigate the ablation study through two parts:

\textbf{RQ2-1} In the vulnerability reachability analysis phase, we propose parameter transfer analysis to eliminate method call paths where vulnerability conditions are unreachable, thereby reducing the generation of unit tests that cannot trigger vulnerabilities.
This part evaluates whether parameter transfer analysis can help improve test validation accuracy.

\textbf{RQ2-2} Well-crafted prompts have the potential to generate higher-quality vulnerability validation unit tests. 
This part aims to validate whether our designed prompt template outperforms the default prompt in terms of performance.
\end{itemize}

These research questions guide our evaluation to assess the effectiveness of \appname in analyzing library vulnerabilities.

\subsection{Experimental Setup}
\noindent \textbf{Dataset.} We collect 30 TPL vulnerabilities from publicly disclosed vulnerability databases such as CVE, covering various categories including XML data injection, remote code execution, and denial of service.
For each collected vulnerability type, we collect client projects from GitHub as our experiment targets.
We follow several criteria throughout the collection process: 
first, each vulnerability needs to have a corresponding publicly available Proof of Concept (PoC) code to ensure the confirmation of the vulnerability's existence;
second, client projects are written in Java, managed under the Maven~\cite{maven} system, and are capable of being compiled and executed successfully;
lastly, client projects depend on versions of TPL vulnerabilities, and based on this, we use call graph analysis to filter out method calls on vulnerable code.
We exclude projects showing package dependencies on TPLs but do not utilize the vulnerable code.
Finally, from the remaining project results, we select the first two client projects as experimental projects corresponding to each vulnerability type.
These CVEs, TPLs, and client project lists, among other detailed data, are publicly available at~\cite{vulEUT_Datasets}.

\noindent \textbf{Baselines.} We compare our approach with \textsc{TRANSFER}~\cite{kang2022test}, and \textsc{VESTA}~\cite{chen2023exploiting}. 
\textsc{TRANSFER} generates tests for exploiting library vulnerabilities based on code behavior observed during vulnerability observation testing.  
Besides, we compare our approach with the state-of-the-art tool \textsc{VESTA}, which generates tests for exploiting library vulnerabilities based on PoC migration.
Both tools provide available replication packages, which we utilize in our experimental setup for comparison.

\noindent\textbf{LLM.} 
We select ChatGPT~\cite{chatgpt} as the default LLM, and we integrate it into our approach using the official ChatGPT interface~\cite{gpt3_5}. 
In terms of usage, we employ the default \texttt{gpt-3.5-turbo} model, known for its stable performance, and the model parameters are set based on the default configurations for optimal code generation practices.

For each client project, we collect the analyzed method's call paths.
We use the \textit{JUnit}~\cite{junit} framework for each call path to generate two unit tests. 
Then, we deploy the generated unit test code in the client project's test directory.
Furthermore, our approach provides test utility classes capable of automatically confirming whether vulnerabilities can be triggered. 
These utility classes are also placed in the test directory.
If a unit test can compile and pass when executed, it is marked as success; otherwise, it is marked as a failure.

% \gao{We conduct experiments on a Mac Pro M1 device with a 3.5GHz processor and 16GB of RAM. 
% Based on the results of Kang et al.'s experiments, we set a 60-second test generation time budget for both methods.}

\subsection{RQ1: VulEUT Effectiveness}
\textbf{Test Generation.} The results of vulnerability exploitation unit tests generated for all 20 TPLs are summarized in Table~\ref{tab:tests}. 
Since we generate two unit tests for each method call path for validation, 292 unit tests are generated across all 146 reachable vulnerability call paths.

\begin{table}
\caption{Results of Unit Test Generation. TPL: Third-Party Libraries, \#Tests: Total tests generated, \#CT: Total tests successfully compiled, \#TT: Total tests successfully confirmed \#CPR: Compilation Pass Rate for Tests.}
 
\centering
\begin{tabular}{l c c c c c}
\toprule
\textbf{TPL} & 
\textbf{\#Tests} & 
\textbf{\#CT} & 
\textbf{\#TT} & 
\textbf{\#CPR} &
\textbf{Accuracy(\%)} \\
\midrule

Apache Codec & 
12 & 
12 & 
12 & 
100 &
100 \\

\hline

Apache Lang & 
32 & 
24 & 
24 & 
75 &
75 \\

\hline

Apache Text & 
10 & 
6 & 
6 & 
60 &
60 \\

\hline

Json-smart & 
10 & 
10 & 
6  & 
100 &
60 \\

\hline

JSON & 
4 & 
3 & 
3 & 
75 &
75 \\

\hline

XStream & 
84 & 
84 & 
63 & 
100 &
75 \\

\hline

Apache.poi & 
4 & 
4 & 
4 & 
100 &
100 \\

\hline

Zip4j & 
8 & 
6 & 
6 & 
75 &
75 \\

\hline

Apache IO & 
32 & 
32 & 
28 & 
100 &
87.5 \\

\hline

Apache PDFBox & 
4 & 
4 & 
4 & 
100 &
100 \\

\hline

Apache Tika & 
4 & 
2 & 
2 & 
50 &
50 \\

\hline

Apache Compress & 
30 & 
20 & 
20 & 
66.67 &
66.67 \\

\hline

Httpclient & 
6 & 
5 & 
5 & 
83.33 &
83.33 \\

\hline

Jsoup & 
14 & 
14 & 
12 & 
100 &
85.71 \\

\hline

Log4j2 & 
32 & 
32 & 
32 & 
100 &
100 \\

\hline

Bouncy Castle &
4 & 
4 & 
4 & 
100 &
100 \\
\hline

\textbf{Total} & 
\textbf{292} & 
\textbf{260} & 
\textbf{229} & 
\textbf{89.04} & 
\textbf{78.42} \\

\bottomrule
\end{tabular}
\label{tab:tests}
% \vspace{-0.4cm}
\end{table}

We deploy each unit test in the respective client project for compilation and execution to confirm the results. 
As shown in Table~\ref{tab:tests}, out of all the generated tests, 260 unit tests are successfully compiled, accounting for 89.04\%, and among them, 229 tests passed confirmation. 
A test is considered to confirm successful if it satisfies the conditions for triggering the vulnerability by utilizing the vulnerability-triggering parameters, reaching the vulnerability code in the TPL, and successfully exploit the vulnerability. 
This results in an accuracy rate of 78.42\% across all 292 tests, indicating that the \appname can not only generate syntactically correct unit tests but also demonstrate effectiveness in confirming vulnerability exploitation.

\begin{table*}
\centering
\caption{Vulnerability Confirmation Results - Each vulnerability corresponds to two client projects. Successful confirmation is indicated by \ding{51}, unsuccessful confirmation by \ding{55}, and the absence of test generation is represented as \textbf{--}.}
{\smaller
\begin{tabular}{c c c c c c c}
\toprule
\textbf{Category} & 
\textbf{TPL} & 
\textbf{CVE} & 
\textbf{Trigger Condition} & 
\textbf{\appname} & 
\textbf{TRANSFER} &
\textbf{VESTA} \\
\midrule

Base64 & 
Apache Codec & 
\makecell[c]{CODEC-263\\CODEC-270} & 
\makecell[c]{Wrong Behavior\\Wrong Behavior} & 
\makecell[c]{\ding{51} ~ \ding{51} \\ \ding{51} ~ \ding{51}} & 
\makecell[c]{\textbf{--} ~ \textbf{--} \\ \ding{51} ~ \ding{51}} & 
\makecell[c]{\ding{51} ~ \ding{55} \\ \ding{51} ~ \ding{51}} \\

\hline

Number & 
Apache Lang & 
\makecell[c]{LANG-1484\\LANG-1645\\LANG-1385} & 
\makecell[c]{Wrong Behavior\\Wrong Behavior\\Wrong Behavior} & 
\makecell[c]{\ding{51} ~ \ding{51} \\ \ding{51} ~ \ding{51} \\ \ding{51} ~ \ding{55}} & 
\makecell[c]{\textbf{--} ~ \textbf{--} \\ \ding{55} ~ \ding{55} \\ \ding{55} ~ \ding{55}} & 
\makecell[c]{\ding{51} ~ \ding{55} \\ \ding{51} ~ \ding{51} \\ \ding{51} ~ \ding{51}} \\
\hline

String & 
Apache Text & 
\makecell[c]{TEXT-215\\CVE-2022-42889} & 
\makecell[c]{Wrong Behavior\\Remote Code Execution} & 
\makecell[c]{\ding{55} ~ \ding{55} \\ \ding{51} ~ \ding{51}} & 
\makecell[c]{\textbf{--} ~ \textbf{--} \\ \ding{55} ~ \ding{55}} &
\makecell[c]{\ding{51} ~ \ding{51} \\ \ding{51} ~ \ding{55}} \\

\hline

JSON & 
\makecell[c]{Json-smart\\ \\JSON\\Jackson-databind} & 
\makecell[c]{CVE-2023-1370\\CVE-2021-27568\\CVE-2022-45688\\CVE-2019-14540} & 
\makecell[c]{Stack Overflow\\Uncatch Exception\\Uncatch Exception\\Remote Code Execution} & 
\makecell[c]{\ding{51} ~ \ding{51} \\ \ding{51} ~ \ding{51} \\ \ding{51} ~ \ding{51} \\ \ding{55} ~ \ding{55} \\} & 
\makecell[c]{\ding{55} ~ \ding{55} \\ \ding{51} ~ \ding{55} \\ \ding{51} ~ \ding{55} \\ \ding{55} ~ \ding{55} \\} &
\makecell[c]{\ding{51} ~ \ding{55} \\ \ding{51} ~ \ding{55} \\ \ding{51} ~ \ding{51} \\ \ding{55} ~ \ding{55} \\} \\

\hline

XML & 
XStream & 
\makecell[c]{CVE-2017-7957\\CVE-2021-39144\\CVE-2021-21341\\CVE-2022-41966\\CVE-2020-26217\\CVE-2020-26258} & 
\makecell[c]{Uncatch Exception\\Remote Code Execution\\Infinted Loop\\Stack Overflow\\Remote Code Execution\\Remote Code Execution} & 
\makecell[c]{\ding{51} ~ \ding{51} \\ \ding{51} ~ \ding{51} \\\ding{51} ~ \ding{51} \\ \ding{51} ~ \ding{51} \\ \ding{51} ~ \ding{51} \\ \ding{51} ~ \ding{51}} & 
\makecell[c]{\ding{51} ~ \ding{51} \\ \textbf{--} ~ \textbf{--} \\ \textbf{--} ~ \textbf{--} \\ \ding{51} ~ \ding{51} \\ \ding{55} ~ \ding{55} \\ \ding{55} ~ \ding{55}} &
\makecell[c]{\ding{51} ~ \ding{51} \\ \ding{51} ~ \ding{51} \\ \ding{51} ~ \ding{51} \\ \ding{51} ~ \ding{51} \\ \ding{51} ~ \ding{51} \\ \ding{51} ~ \ding{51}} \\

\hline

File & 
\makecell[c]{Apache Tika\\Apache.poi\\Zip4j\\Apache IO\\ \\Apache PDFBox} & 
\makecell[c]{CVE-2019-10094\\CVE-2019-12415\\CVE-2022-24615\\IO-611\\CVE-2021-29425\\CVE-2021-31812} & 
\makecell[c]{Infinted Loop\\XXE Injection\\Uncatch Exception\\Path Traversal\\Path Traversal\\Stack Overflow} & 
\makecell[c]{\ding{51} ~ \ding{55} \\ \ding{51} ~ \ding{51} \\ \ding{51} ~ \ding{51} \\ \ding{51} ~ \ding{51} \\ \ding{51} ~ \ding{51} \\ \ding{51} ~ \ding{51}} & 
\makecell[c]{\ding{51} ~ \ding{55} \\ \textbf{--} ~ \textbf{--} \\ \textbf{--} ~ \textbf{--} \\ \ding{55} ~ \ding{55} \\ \ding{55} ~ \ding{55} \\ \textbf{--} ~ \textbf{--}} &
\makecell[c]{\ding{55} ~ \ding{55} \\ \ding{51} ~ \ding{51} \\ \ding{51} ~ \ding{55} \\ \ding{51} ~ \ding{51} \\ \ding{51} ~ \ding{51} \\ \ding{51} ~ \ding{51}} \\ 

\hline

Net & 
Httpclient & 
\makecell[c]{CVE-2020-13956\\HTTPCLIENT-1803} & 
\makecell[c]{Cross-site Scripting\\Wrong Behavior} & 
\makecell[c]{\ding{51} ~ \ding{55} \\ \ding{55} ~ \ding{51}} & 
\makecell[c]{\ding{55} ~ \ding{55} \\ \ding{55} ~ \ding{55}} &
\makecell[c]{\ding{51} ~ \ding{51} \\ \ding{55} ~ \ding{51}} \\

% \makecell[c]{CVE-2021-35516\\CVE-2018-1324} & 
% \makecell[c]{Out of Memory\\Wrong behavior} & 
% \makecell[c]{ } & 
% \makecell[c]{\ding{51} ~ \ding{55} \\ \textbf{--} ~ \textbf{--} } &
% \makecell[c]{\ding{51} ~ \ding{55} \\ \ding{51} ~ \ding{51}} \\

\hline

HTML & 
Jsoup & 
\makecell[c]{CVE-2021-37714} & 
\makecell[c]{Remote Code Execution} & 
\makecell[c]{\ding{51} ~ \ding{51}} & 
\makecell[c]{\ding{55} ~ \ding{55}} &
\makecell[c]{\ding{51} ~ \ding{55}} \\

\hline

Cryptography & 
Bouncy Castle & 
\makecell[c]{CVE-2020-28052} & 
\makecell[c]{Wrong Behavior} & 
\makecell[c]{\ding{51} ~ \ding{51}} & 
\makecell[c]{\ding{51} ~ \ding{51}} &
\makecell[c]{\textbf{--} ~ \textbf{--}} \\

\hline

Log & 
Log4j2 & 
\makecell[c]{CVE-2021-44228\\CVE-2021-45046} & 
\makecell[c]{SQL injection\\Remote Code Execution} & 
\makecell[c]{\ding{51} ~ \ding{51} \\ \ding{51} ~ \ding{51}} & 
\makecell[c]{\ding{55} ~ \ding{55} \\ \ding{55} ~ \ding{55}} &
\makecell[c]{\ding{51} ~ \ding{55} \\ \ding{51} ~ \ding{55}} \\

\hline 

Database & 
Hibernate & 
\makecell[c]{CVE-2019-14900} & 
\makecell[c]{SQL injection} & 
\makecell[c]{\ding{55} ~ \ding{55}} & 
\makecell[c]{\ding{55} ~ \ding{55}} &
\makecell[c]{\ding{55} ~ \ding{55}} \\

\hline

Compress & 
Apache Compress & 
\makecell[c]{CVE-2021-35516\\CVE-2018-1324\\CVE-2023-42503} & 
\makecell[c]{Out of Memory\\Wrong behavior\\Wrong behavior} & 
\makecell[c]{\ding{51} ~ \ding{51} \\ \ding{51} ~ \ding{51} \\ \ding{51} ~ \ding{51}} & 
\makecell[c]{\ding{51} ~ \ding{55} \\ \textbf{--} ~ \textbf{--} \\ \textbf{--} ~ \textbf{--}} &
\makecell[c]{\ding{51} ~ \ding{55} \\ \ding{51} ~ \ding{51} \\ \ding{55} ~ \ding{55}} \\

\hline
Test & 
JUnit & 
\makecell[c]{CVE-2020-15250} & 
\makecell[c]{Improper File Permission} & 
\makecell[c]{\textbf{--} ~ \textbf{--}} & 
\makecell[c]{\ding{55} ~ \ding{55}} &
\makecell[c]{\ding{55} ~ \ding{55}} \\

\hline 

Framework & 
Spring-beans & 
\makecell[c]{CVE-2022-22965} & 
\makecell[c]{Remote Code Execution} & 
\makecell[c]{\textbf{--} ~ \textbf{--}} & 
\makecell[c]{\ding{55} ~ \ding{55}} &
\makecell[c]{\textbf{--} ~ \textbf{--}} \\

\hline

& 
20 TPLs & 
35 Vulnerabilities & 
& 
56/70 $ \mathbf{(24.44\% \uparrow)} $& 
12/70 &
45/70
\\
\bottomrule
\end{tabular}
}
\label{tab:verify}
\vspace{-0.4cm}
\end{table*}

We compare our vulnerability confirmation results with the \textsc{TRANSFER}, as shown in Table \ref{tab:verify}. 
In all completed vulnerability confirmation projects, our approach effectively confirms the presence of vulnerabilities in 56 out of 70 projects. 
In the same dataset, \textsc{TRANSFER} achieves confirmation as intended in only 12 projects. 
The testing generation process for \textsc{TRANSFER} relies on existing vulnerability-fix tests within TPL. 
However, it is worth noting that such tests do not exist in all TPLs. 
For example, in the case of CVE-2022-24615, the \texttt{Zip4j} library lacks vulnerability-fix tests.
Additionally, tests generated based on code behavior are not always entirely consistent with those that can trigger vulnerabilities, especially in scenarios involving file operations like CVE-2019-12415. 
This discrepancy results in generated tests failing to trigger the vulnerability, thereby limiting the effectiveness of \textsc{TRANSFER}.

Regarding \textsc{VESTA}, migrating test generation from PoCs resolves the dependency on TPL vulnerability tests. 
Nevertheless, \textsc{VESTA} is limited by manually defined parameter conversion rules, particularly when these rules are not adequately covered for various parameter types, which can lead to failures and errors in the generated tests.
For example, in CVE-2021-37714, the PoC provides a maliciously compressed HTML file. 
The client method \textit{extract} accepts a string parameter of HTML type, necessitating parsing and character conversion for the compressed file. 
\textsc{VESTA} lacks such parameter conversion rules, resulting in errors in the generated test. 
Additionally, \textsc{VESTA} struggles with handling complex multi-parameter vulnerabilities, such as CVE-2023-42503.
In this case, the client method accepts a file path and two file directory parameters as inputs, but the PoC only provides one malicious file path parameter. 
\textsc{VESTA} fails to handle the other two directory parameters, leading to test failure.

Our \appname leverages the code generation capabilities of LLM by providing descriptions of parameter usage to the model. 
\appname can automatically parse and process these parameter types. 
For example, CVE-2021-37714 generates the corresponding parsing and conversion code for the malicious HTML file, translating it into a string format. 
Additionally, for multi-parameter cases, \appname understands how to construct valid parameter combinations. 
For CVE-2023-42503, it generates the file path for the client target code, enabling the resulting test to run successfully and trigger the vulnerability.

In the experiments, the \appname can complete code analysis and automatically generate prompts for unit tests within 30 seconds, showing its exceptional performance and potential for practical application.
%\begin{figure}
% \vspace{0.3cm}
% \begin{mdframed}[backgroundcolor=gray!7]
% \textbf{Answer to RQ1:}
% \appname can generate unit tests for confirming vulnerability exploitation in client projects.
% It successfully generates 229 verifiable tests from 70 client projects, substantially outperforming the baseline tools (\textsc{TRANSFER} and \textsc{VESTA}), which demonstrates the effectiveness of our approach.
% \end{mdframed}

%\vspace{-5pt} 
%\end{figure}

\subsection{RQ2: Ablation Study}
Next, we conduct an ablation study for the \appname.

\textbf{RQ2-1: Does parameter transfer analysis improve VulEUT performance?} We propose parameter transfer analysis to the method call path analysis and assess its impact on the performance of our approach. 

Experimental results, as shown in Table~\ref{tab:ab1}, reveal that 
combining parameter transfer analysis in test generation achieves an accuracy of 78.42\%.
In contrast, if tests are generated only based on method call paths, this results in a total of 355 generated tests, of which 126 can not pass vulnerability trigger confirmation, with an accuracy rate of only 64.5\%. 
% when generated tests with the combination of parameter transfer analysis, the accuracy reaches 78.42\%.

Similar to the scenario illustrated by the method call path within the \textit{OpenOLAT} project, as shown in Section~\ref{sec:motivation}, this specific call path encompasses invocations of the vulnerable method. 
However, it is notable that the input parameters at the entry point initiate a network request, while the input to the vulnerability method is derived from parsing and processing the result of the network request.
This results in an interruption in the propagation of parameters from the entry point to the vulnerability method. 
Therefore, it cannot directly propagate malicious values and trigger the vulnerability. 
In such cases, this call path should be filtered out.

Through parameter transfer analysis, our approach eliminates the method call path that cannot be passed from the entry to the vulnerability code, thereby avoiding the generation of unit tests that cannot exploit the vulnerability, and thus improves the performance, which shows the effectiveness of our approach in the vulnerability reachability analysis phase.

\begin{table}
\caption{Performance comparison of VulEUT before and after incorporating parameter analysis, using default method call paths with parameter analysis.}

\centering
\resizebox{\linewidth}{!}{
\begin{tabular}{l c c c}
\toprule
\textbf{Variant of the Approach} & 
\textbf{\#Tests} & 
\textbf{\#TT} & 
\textbf{Accuracy} \\
\midrule

\appname (Only Method Paths) & 
355 & 
229 & 
64.5\% \\

\appname (With Parameter Transfer Analysis)  & 
292 & 
229 & 
\textbf{78.42\%} \\

\bottomrule
\end{tabular}}
\label{tab:ab1}
% \vspace{-0.4cm}
\end{table}

\textbf{RQ2-2: Does prompt template design improve the performance of VulEUT?} We design three types of prompts for testing generated prompts: 1) Default prompts, 2) Template prompts with a zero-shot mechanism, and 3) Template prompts with a few-shot mechanism.

\begin{table}
\caption{Performance of VulEUT in generating vulnerability confirmation tests under different variants.}

\centering
\begin{tabular}{l c c c}
\toprule
\textbf{Variant of the Approach} & 
\textbf{\#Tests} & 
\textbf{\#TT} & 

\textbf{Accuracy} \\
\midrule

\appname (Default Prompt) & 
292 & 
138 & 
47.26\% \\

\appname (Zero-Shot Template)  & 
292 & 
185 & 
63.36\% \\

\appname (Few-Shot Template)  & 
292 & 
229 & 
\textbf{78.42\%} \\

\bottomrule
\end{tabular}
\label{tab:ab2}
\vspace{-0.4cm}
\end{table}

% Default prompts only provide a basic outline of the requirements in natural language and rely on ChatGPT~\cite{chatgpt} for generating unit tests. 
Table~\ref{tab:ab2} shows the performance of the \appname under these three prompt design schemes. 
The default prompt is used to describe the test generation task in natural language, providing the signature of the focal method, the class where the method resides, and the testing intent. 
It can be observed that default prompts result in low-quality test generation, with an accuracy of only 47.26\% across the three generated tests. 
However, compared to our designed prompt template, it lacks important information such as Test Oracle and Vulnerable Method (see Table~\ref{tab:prompt-template}). 
This omission may result in generated tests lacking accuracy and completeness in functionality.

The employing of template prompts with zero-shot learning mechanisms significantly improves the test quality, with an accuracy of 63.36\%.
We illustrate this result through practical examples. 
Figure~\ref{fig:prompt-example} displays two generated unit tests, where Test (a) is a test generated under the default prompt, while Test (b) is a test generated under the zero-shot template.
It can be seen that Test (a) has two evident issues. 
First, it contains code with ambiguous semantics, such as the utilization of \textit{YourClass}, which is due to the lack of sufficient method parameter information provided in the prompt to LLM, leading to ambiguous code. 
This issue can render the generated test code inexecutable, given its inability to compile.
Second, Test (a) lacks a precise test oracle; it merely expects an exception to be thrown during execution.
However, the mere occurrence of an exception does not necessarily indicate the successful triggering of a vulnerability, rendering it imprecise.

\begin{figure}
   \begin{lstlisting}[language=Java,xleftmargin=1em,title=(a) Example of Test Generated by VulEUT (Default Prompt),linewidth=\linewidth]
 @Test(expected = Exception.class)
 public void testVulnerabilityTriggered() {
  // Input data
  String xml = "<void>"; 
  // Replace with the appropriate class for your use case
  Class<?> clazz = YourClass.class; 
  // Call the vulnerable method
  XmlUtil.xml2Obj(xml, clazz); 
}
    \end{lstlisting}
    \begin{lstlisting}[language=Java,xleftmargin=1em,title=(b) Example of Test Generated by VulEUT (Zero-Shot),linewidth=\linewidth]
 @Test 
 public void testXml2ObjCallVulnerabilityMethod(){
  String input = "<void>";
  //Set up an interceptor to detect calls to the focal method
  MethodCallInterceptor.interceptor(
        com.thoughtworks.xstream.XStream.class, "fromXML", new Object[]{input});
  XmlUtil.xml2Obj(input, Object.class);
  // Verify if a call to the focal method
  assertTrue(MethodCallInterceptor.isTriggered());
  assertTrue(MethodCallInterceptor.isConditionMet());
}
    \end{lstlisting}
\caption{Examples of VulEUT test generation under different prompt design schemes.}
\label{fig:prompt-example}
\vspace{-0.4cm}
\end{figure}

In contrast, Test (b) benefits from our prompt template, effectively resolving the issues witnessed in Test (a).
First, our template supplies comprehensive parameter information, avoiding the LLM's reliance on ambiguous pseudo-code. 
Second, our template requires the utilization of vulnerability confirmation components provided by our approach, ensuring that LLM employs the correct assertion.
Test (b) not only compiles successfully but also accurately confirms whether the vulnerability is triggered. 
Nevertheless, Test (b) also has an issue where the focal method may encounter an exception during execution, leading to the test exiting abnormally.
This would render assertion unusable in determining whether the TPL vulnerability is triggered.

Our expectation is that the focal method is wrapped in a try-catch block, ensuring that it does not exit directly when an exception occurs during test execution. 
More importantly, our approach ensures the intended execution of our assertions, thereby facilitating vulnerability confirmation.
Hence, within the framework of our few-shot learning mechanism, we provide LLM with five real-world examples of focal methods wrapped in try-catch blocks and proceed with the test generation task. 
In this mode, LLM learns to structure tests from these examples, enhancing \appname's accuracy.

% \vspace{0.3cm}
% \begin{mdframed}[backgroundcolor=gray!7]
% \textbf{Answer to RQ2:}
% Method call path analysis, especially with parameter transfer analysis, significantly improved \appname performance.
% Prompt templates, particularly when combined with few-shot examples, considerably enhanced \appname performance, demonstrating the effectiveness of this approach.
% \end{mdframed}

\section{Discussion}
\label{sec:discussion}
\subsection{Strengths of \appname}

Compared to data flow graphs, PTG focuses more on analyzing the parameter transfer process rather than merely describing dependencies between data. 
PTG provides more detailed information, including code statements for parameter changes (such as assignment statements, function computations, and type conversion statements) and syntax trees. 
Combined with the parameter propagation rules (Table~\ref{tab:rule-desc}), it helps determine the reachability of vulnerable conditions in paths.
For example, in CVE-2017-7957, using only method call path analysis, LLM generated a unit test for the \texttt{getDiscovery} method. 
Although LLM generates the \texttt{discoveryUrl} parameter value, such a test fails to transfer malicious XML values specific-crafted and capable of triggering the vulnerability. 
Therefore, this unit test is redundant.
Our approach avoids generating unit tests by LLMs for unreachable paths with parameters, ensuring that the generated tests can trigger the vulnerability through malicious input.

\subsection{LLM-Based Vulnerability Confirmation}
Our tool \appname demonstrates the effectiveness of LLM in generating unit tests for vulnerability confirmation. 
By combining code analysis and prompt template design, we not only mitigate LLM hallucination issues but also demonstrate its generalization capability.
Regarding hallucination issues, a current solution is based on information retrieval methods, where LLMs are initially tasked with searching and generating code from specified code repositories. 
Our approach is similar to this, where through designed prompt templates, we provide necessary information for each component of the test, enabling LLMs to assemble and generate complete tests. 
Additionally, our few-shot learning further regularizes the language model generation process, assisting in mitigating hallucination issues. 

Our tool also utilizes the generalization capability of LLMs, as the parameter types of entry methods in client projects may differ from those in vulnerability APIs within TPLs, resulting in diversity during test generation. 
For example, in CVE-2021-29425, the \texttt{normalize} API in the \texttt{commons.io} library accepts a string parameter. 
When an incorrect input string (e.g., \texttt{"//../foo"}) is used, it erroneously grants access to files in the parent directory. 
In the \textit{velocity-engine} project, although there exists a call path to \texttt{normalize}, besides string input, there is also a character encoding parameter. 
Due to different encoding values leading to different processing paths, \appname generates various encoding values such as \texttt{UTF-8}, \texttt{ISO-8859}, and \texttt{ASCII}, demonstrating diversity.

\subsection{Threats to Validity}

% \subsection{Limitations}
% One limitation of \appname is that when dealing with large-scale functions, the generated unit tests may encounter issues.   
% In our experiments, we identify an entry function in the \textit{Jsoup} project that exceeds 150 lines of code, resulting in an excessively long prompt. In such cases, the tests may exhibit syntax errors.
% The main issue causing this problem is that the LLM has trouble handling exceptionally lengthy prompts, which can lead to inadvertent omissions of specific information and, an incomplete understanding of the problem.   
% As a result, the LLM fails to capture all essential details from our prompt template, resulting in issues in the generated unit test code.
Threats to external validity relate to the ability of our approach in generating unit tests applicable to a broader range of projects and various vulnerability types. 
To address this issue, we employ a random sampling method to collect real projects from various categories on GitHub. We aim to assess its applicability in domains beyond those we have evaluated, although there may still be performance variations in areas we have not explored yet.

Threats to internal validity pertain to the source code analysis method we employ. 
We use a class-level declaration-based approach to obtain call graphs within projects, which might introduce errors in certain cases. We actively explore more accurate analysis methods to mitigate this potential source of error. Previous studies ~\cite{callgraph, graph} also perform the same approach to obtain the call graphs.

% Another threat relates to the use of the ChatGPT model. 
% We utilize the default parameters provided by the official version. 
% Fine-tuning the model with different parameters could potentially enhance the performance of test generation, and we plan to investigate it in our future works. Additionally, regarding the design of prompt templates, we follow~\cite{pearce2023examining} to design the prompts, and we strive to make them as generic as possible. 
% However, slight performance variations may occur in projects from different domains.
\section{RELATED WORK}
\label{sec:relatedwork}

\noindent \textbf{Vulnerability Reachability.}
Security issues related to TPLs have been a prominent focus of research~\cite{duan2020towards, huang2022characterizing,decan2019empirical,zimmermann2019small,alfadel2023empirical,liu2022demystifying,chen2023identifying,zhang2023compatible}. 
% Ponta et al.~\cite{ponta2018beyond} combined static and dynamic analysis tools to detect the reachability of vulnerable code, which overcomes their mutual limitations.
Mir et al.~\cite{mir2023effect} investigated the propagation patterns of vulnerabilities in the Maven ecosystem. They analyzed the distribution of vulnerabilities and the differences in vulnerability propagation at the package and method levels. Their approach relied on project dependencies, program call graphs, and conducted reachability analysis. 
The study emphasized the critical importance of considering vulnerability reachability at various granularity levels, particularly for security-focused applications. They suggested that graph-based analysis methods hold promise for future research. 
% Wu et al.~\cite{wu2023understanding} conducted an empirical study on the impact of TPL vulnerabilities on software projects, analyzing CVEs, libraries, and project data. They examined vulnerability reachability and path complexity using call and control flow graphs but did not actively trigger these vulnerabilities in practice.
Kang et al. proposed \textsc{TRANSFER}~\cite{kang2022test} to analyze the call relation of the methods in the client project and implied existing tools to generate a test case similar to the vulnerable witness test, which is considered a test case that could trigger the vulnerability. 
By executing the generated test case and manually checking the trigger condition of the vulnerability, \textsc{TRANSFER} accessed the reachability of the vulnerability precisely.  
Different from the aforementioned studies, our approach not only analyzes the reachability of vulnerability parameters between the client entry point and TPL vulnerability code but also generates unit test cases to confirm the presence of vulnerabilities. 
Our method checks the reachability of the vulnerabilities by generating a test case that covers the vulnerable method, adding an assertion to check the trigger of the vulnerability, and avoiding false alerts.

\section{Conclusion and Future Work}
\label{sec:conclusion}
This paper presents an approach called \appname, aimed to assess the exploitability of TPL vulnerabilities in client projects. 
Our approach combines source code analysis with LLM's test generation capabilities to generate unit tests that can trigger vulnerabilities in client projects. 
Compared to existing technologies, \appname can more accurately demonstrate the exploitability of TPL vulnerabilities within client projects. 
It shows seven times more client projects where vulnerabilities can be exploited compared to the \textsc{TRANSFER}, confirming the effectiveness of the approach across various domains and vulnerability types in open-source projects.
In the future, we intend to expand our research by conducting additional experiments on a wider range of datasets and various types of vulnerabilities. Additionally, we plan to explore how the performance of our approach compares across different categories of open-source Large Language Models.

\section*{Acknowledgment}
This work was supported by the National Key R\&D Program of China (No. 2024YFB4506400) and Ningbo Natural Science Foundation (No. 2023J292).

\balance
% \begin{thebibliography}{1}
\bibliographystyle{IEEEtran}
\bibliography{main}
% \end{thebibliography}
\end{document}